\documentstyle[times,namedreferences]{spackap}
\begin{opening}
\title{ Low Frequency Flickering of TT Arietis:Hard and Soft X-ray Emission Region}
\author{Altan \surname{Baykal}}
\author{\"Umit  \surname{KIzIlo\u{g}lu}}
\institute{Physics Department, Middle East Technical University, Ankara 06531, Turkey }
\date{}
\end{opening}
\begin{document}

\begin{abstract}
 Using archival ASCA observations of TT Arietis, X-ray energy spectra 
 and power spectra of the intensity time series   
 are presented for the first time. The energy 
  spectra  are well-fitted by a two  continuum  plasma emission model 
with temperatures $\sim  1$ keV and $\sim 10$ keV.    
 A coherent feature at $\sim  0.643$ mHz 
 appeared in the power spectra during the observation.
   
\end{abstract}

\section{Introduction}

 TT Arietis has been categorized as a nova-like variable from 
photometric observations (Smak \& Stepien 1969, Cowley et al., 1975). 
The system's spectroscopic orbital period is found to be 0.13755114(13)$^{d}$ 
(Thorstensen et al., 1985) from measurements of the radial velocities 
of emission
lines. This is longer than the photometric 
period of 0.1329$^{d}$  (Tremko 1992).  The beat frequency of the photometric 
 and spectroscopic orbital periods is $\sim 0.25$ d$^{-1} $ which is 
 very similar to the intermediate polar TV Col (Hellier et al. 1991). 
This led Jameson et al. (1982) to the suggestion that TT Ari is an 
intermediate polar.  
 The optical brightness of  
TT Ari varies irregularly on long time scales. It can remain at 
visual magnitudes $\sim 10 $ (high state) for many years, 
interrupted irregularly by low states with 
visual magnitudes $\sim 16 $  
which last less than a year (Shafter et al. 1985). In the high state, 
TT Ari has shown flickering activity (or quasi-periodic oscillations) 
  with periods between 
14 min and 27 min (Semeniuk et al., 1987; 
Hollander \& van Paradijs 1992).  

TT Ari was found to be a hard X-ray source by the
Einstein satellite (Cordova et al., 1981).
 Jensen et al. (1983) investigated the correlation of
X-ray and optical flickering activity
 using simultaneous observations
with the Einstein X-ray observatory and the Mount Wilson Optical Telescope.
They found that the flickering activity of X-rays
at $\sim 17$ min ($\sim 1$ mHz) is delayed by $\sim$ 1 min
with respect to the optical flickering activity. They proposed that the 
optical flickering in TT Ari is produced in the inner
accretion disk, and a fraction of the energy is transported to a
different region where the X-ray flickering is produced. The 1 min time
delay between the optical and X-ray flickering would be the time
required to transport energy from the optical flickering region
to the X-ray emitting region (Jensen et al., 1983). 

 In the present work, we analyze archival ASCA data to construct the 
 X-ray egergy spectra and power spectra of TT Ari. 
Section 2 describes the observations. Section 3 describes the 
X-ray spectra and X-ray timing.

 \section{Observations}

 TT Ari was observed with ASCA on January 20 to 21, 1994   
 with an effective exposure time $\sim 14$~ksec.                 
The ASCA instrumentation (Tanaka et al. 1994) consists of four imaging 
telescopes, each with a dedicated spectrometer. There are two solid-state
  imaging spectrometers (SIS), each consisting of 4 CCD chips, giving 
an energy resolution of 60-120 eV across the 0.4-10 keV band. 
The SIS with 2-CCD mode
 and GIS data were taken with a time resolution 8 sec and 62.5 msec 
respectively.    
Two gas scintillation proportional counter imaging 
spectrometers, GIS, have an energy resolution of 200-600 eV over the 0.8-10 
keV band. Data were extracted within a $\sim 4$ arcmin radius region 
for each SIS and within a $\sim 6$ arcmin region for each 
GIS. The typical mean count rates of TT Ari, for the SIS and GIS are
 $\sim  0.59$ and $\sim 0.48$ count sec$^{-1}$, 
respectively. 
 Standard cleaning for ASCA data was applied to eliminate X-ray
 contamination from the bright Earth, effects due to high particle background, 
and hot flickering SIS pixels. The reduction of the ASCA archival data and 
the 
correction of the photon arrival times to the Solar System 
barycenter was performed using 
 XSPEC, XRONOS, XIMAGE and XSELECT softwares. In Fig. 1, we present the 
light curve of the observation.      
%----------------------------------------------------------- Spec_1
\begin{figure}
\vspace{7.5cm}
\includegraphics{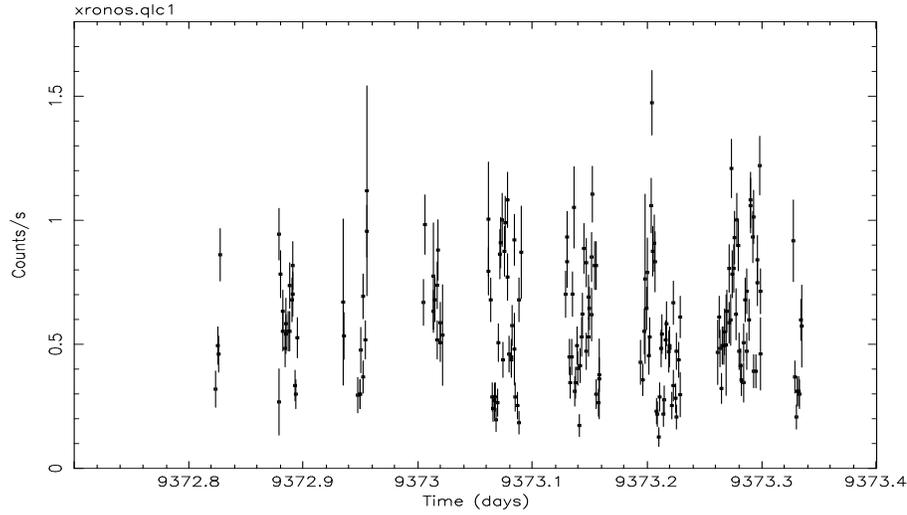}
      \caption{The light curve of TT Ari with SIS0 count rates binned 
               in intervals of 86 sec. The times indicates MJD 40000+.
              }
    \end{figure}

\section{Results}
\subsection{X-Ray Spectra}

  The large number of counts obtained from ASCA observations 
 and the broad energy range (0.5-10 keV) 
 allows for fits using various spectral models.
 The spectrum was fitted with a two component Raymond Smith (1977) 
emission model as expected from radiatively cooling shock regions. 
%The results are displayed in Table 1.  
 The measured unabsorbed flux during the observation was  
$\sim 9.74 \times 10 ^{-12}$~erg~sec$^{-1}$~cm$^{-2}$ in the band 
of 0.5-10 keV. 

 In order to see the lower energy part of the spectrum ($<$ 2 keV)    
more clearly and to estimate the column density N$_{H}$ more accurately,  
ROSAT archival observations  of TT Ari were extracted and a two component
Raymond-Smith model was fitted to combined ASCA and ROSAT data.
Our preliminary analysis showed that  both data 
 sets have similar spectra.  
 In these fits, we assume that the spectral parameters of 
ASCA and ROSAT observations were the same  
 up to a normalization factor. Results of the fit, shown in Table 1, 
imply that the ASCA flux was $1.1 \pm 0.02 $ of the ROSAT flux.
 The column density of N$_{H} \sim (4.13 \pm 0.11) \times 10^{20}$~cm$^{-2}$ 
obtained from the joint fit of the ROSAT and ASCA data sets is consistent 
with the previously 
deduced value from the ROSAT observation alone (Baykal et al., 1995). 
The two component Raymond Smith model fits the data well
 with $\chi^{2}_{\nu} \sim 1.03$, giving  the plasma temperatures of 
about $\sim  1$~keV and $\sim 10$~keV, for the two components.
  
 Keeping in mind that the spectral characteristics may change in time and
 while the column density probably remains constant; the column density 
deduced from the joint ROSAT-PSPC
and ASCA-SIS0+GIS2 data was  used for further spectral work with the ASCA
observations. 
In Fig. 2, we present the energy spectra fits with the  two component model.
Only SIS0+GIS2 data are used for clarity.  
\begin{figure}
\vspace{9cm}
\includegraphics{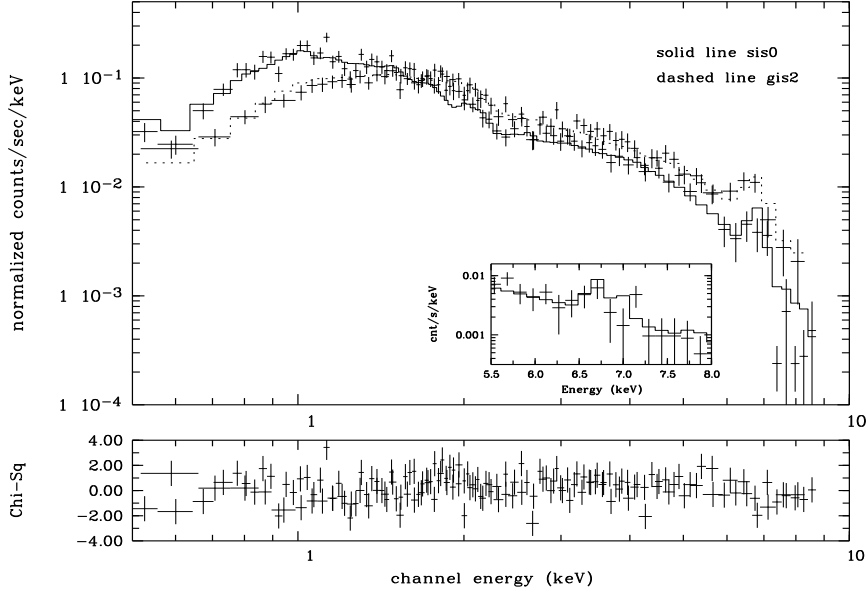}
      \caption{The average ASCA/ SIS0, GIS2 spectra of TT Ari.
            The upper panel shows the fit of the two component
            Raymond-Smith model. Data is binned for 5$\sigma$ confidence.
            The lower panel is the $\chi^{2}$ residuals from the fit.
            Inset shows 6.67 keV feature, binned for 3$\sigma$ confidence,
             together with the model fitted to it.
              }
%       \label{Fig-1}
\end{figure}

Table 1 gives
the spectral parameters for the best fit to the SIS0+GIS2 data 
to compare with the joint fit and 
two component  Raymond Smith model to  SIS0+SIS1+GIS2+GIS3 detectors. 
\begin{table}
\caption{ Spectral Fit Parameters for Two Temperature Models $^{a}$}
\begin{tabular}{l c c c c c c }
\hline
 model  & kT$_{1}$(keV) & kT$_{2}$(keV)
        & A$_{1}(10^{-4}$)$^{f}$ & A$_{2}(10^{-3}$)$^{f}$
        & N$_{H}(10^{20}$cm$^{-2}$)  & $\chi^{2}_{\nu} $ \\
\noalign{\smallskip}
\hline
R-S $^{b}$  &  $1.03\pm0.05 $    &  $10.99\pm1.17 $
            &  $4.92\pm1.36 $    &  $6.83\pm0.14 $
            &  $4.13\pm0.11 $    &  1.03 \\
R-S $^{c}$  &  $0.99\pm0.12 $    &  $9.29\pm1.02 $
            &  $3.70\pm2.22 $    &  $7.16\pm0.19 $
            &  4.13 fixed        &  0.96 \\
R-S $^{d}$  &  $0.97\pm0.08 $    &  $8.49\pm0.63 $
            &  $3.67\pm1.51 $    &  $7.26\pm0.13 $
            &  4.13 fixed        &  1.36 \\
Meka $^{e}$ &  $0.76\pm0.05 $    &  $8.10\pm0.49 $
            &  $3.37\pm0.91 $    &  $7.32\pm0.09 $
            &  4.13 fixed        &  1.48\\
\noalign{\smallskip}
\hline
\end{tabular}

~~$^{a}$~Spectral fits were performed using
 the program XSPEC.\\
~~$^{b,c,d}$~Raymond$-$Smith model for PSPC+SIS1+GIS2, SIS1+GIS2  and
SIS0+SIS1+GIS2+GIS3 detectors respectively. \\
~~$^{e}$~Mewe$-$Kaastra model for SIS0+SIS1+GIS2+GIS3 detectors. \\
~~$^{f}$~~A$_{1}$ and  A$_{2}$ are the emission measures in units of
   $10^{-14} / (4 \pi D^{2}) \int ~~n^{2} dV$, where $D$ is the distance \\
   to the source (cm) and n is the electron density (cm$^{-3}$). \\

\end{table}

Spectral results from a two temperature Mewe-Kaastra model are also given.
This model gives slightly lower plasma temperatures, but it does not
improve the goodness of fit.
 
The residuals  do not indicate an extra line feature at 
$ 6.67 \pm 0.03$~keV. Inset in Fig. 2 shows data and the model fitted 
to it. To check whether a line feature component is evident for 
 6.43 keV which may indicate the presence of of emission from
 the white dwarf surface, a variable Raymond Smith 
model with zero Fe abundance and a Gaussian line feature is tested. 
The data is again well fitted with a Gaussian line at 
$ 6.67 \pm 0.03$~keV of a one sigma width  $ 68 \pm  39$~eV and a line flux  
$(2.42 \pm 0.10) \times 10^{-5}$~photons~cm$^{-2}$sec$^{-1} $.   

 In order to see the spectral modulation over the orbit, hardness 
ratios  (2-10/0.2-2 keV) are plotted in Fig. 3. The data are folded  
 at the spectroscopic orbital period P=0.13755114(13)$^{d}$ with 
 an epoch JD= 2,443,729.0175 (Thorstensen et al., 1985). 
  As seen from the Fig. 3, there is no significant  
 variation of the hardness ratio.  
 The binary orbital phases 0.75-0.9 were softer in ROSAT observations 
 (see Baykal et al., 1995). However, as  these orbital phases are occulted 
in the ASCA observation  due to the satellite orbit around the earth,so 
we can not address this issue with the ASCA data. 

%-----------------------------------------------------------
\begin{figure}
\vspace{7cm}
\includegraphics{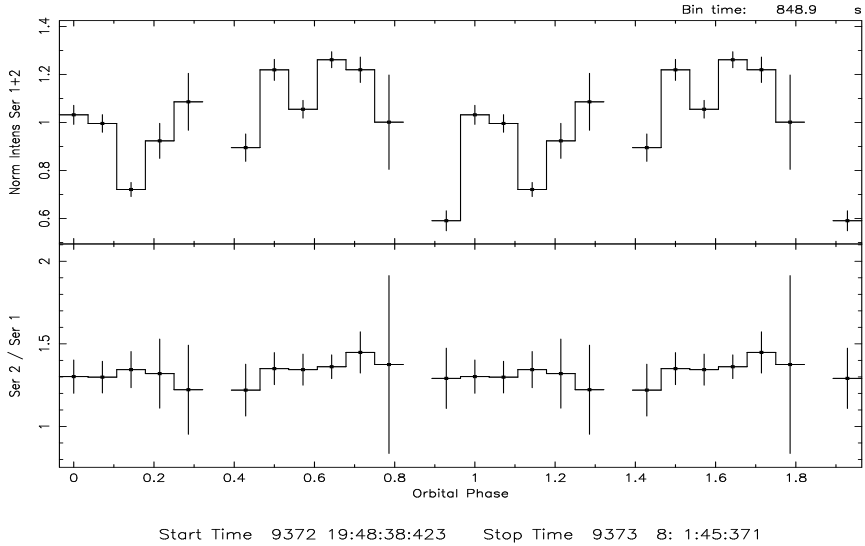}
      \caption{The upper panel is the normalized light curve folded
            at the spectroscopic orbital period P=0.13755114(13)$^{d}$ with
            the epoch JD= 2443729.0175 (Thorstensen et al., 1985).
            The lower panel is the hardness ratio (2-10 keV/0.2-2 keV) of the
            light curve.
              }
%       \label{Fig-1}
\end{figure}
\subsection{X-Ray Timing}

 The ASCA observations were interrupted by a number of gaps.
 In order to examine the high frequency oscillations we calculated
Fast Fourier transforms for each observation window
and then averaged them.

%----------------------------------------------------------- Spec_1
\begin{figure}
\vspace{6.5cm}
\includegraphics{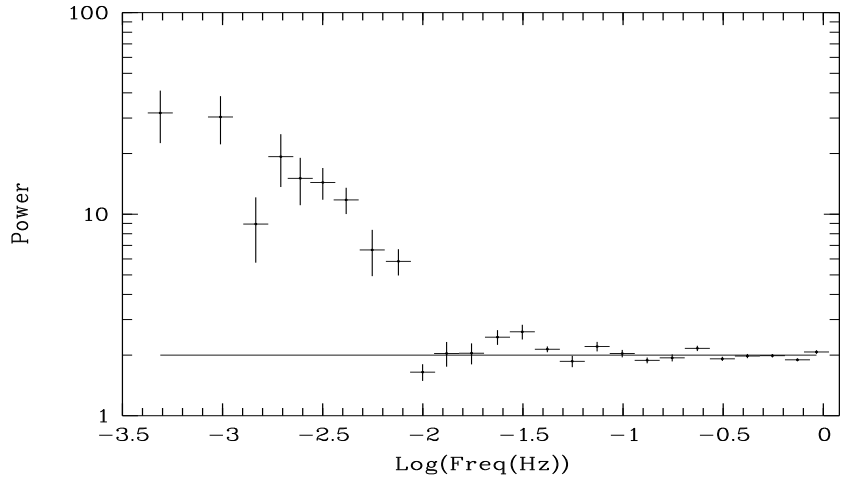}
      \caption{The average power spectrum of TT Ari. Data were binned
            0.5 sec for each GIS2 and GIS3 detectors. The power
            spectra were averaged
            for each observational window and presented by
            log-log scale (horizontal line represents
            the counting statistics noise).
            The windows were of length
            2048 seconds.
              }
%       \label{Fig-1}
\end{figure}

  The length of each stretch was chosen as 2048 sec with 4096 bins.
The average power density spectrum is normalized according to
Leahy et al. (1983) and rebinned logarithmically
in frequency (see Fig. 4). In the power
spectrum, the counting statistics noise (or white noise) level
is around 2 since there are 2 degrees of freedom
for each power estimate and each estimate obeys
a $\chi^{2}$ distribution (see van der Klis 1989).
 For frequencies higher than 10 mHz, the variability of
TT Ari dissolves into the noise level
which is consistent with the ROSAT observations (Baykal et al. 1995).
For frequencies lower
 than 10 mHz, a red noise component (or flicker noise) exists.

In order to examine
 the flickering activity and the oscillations at lower frequencies,
discrete power spectra (Deeming 1975) were estimated and normalized
according to Leahy et al. (1983).

%----------------------------------------------------------- Spec_1
\begin{figure}
\vspace{8cm}
\includegraphics{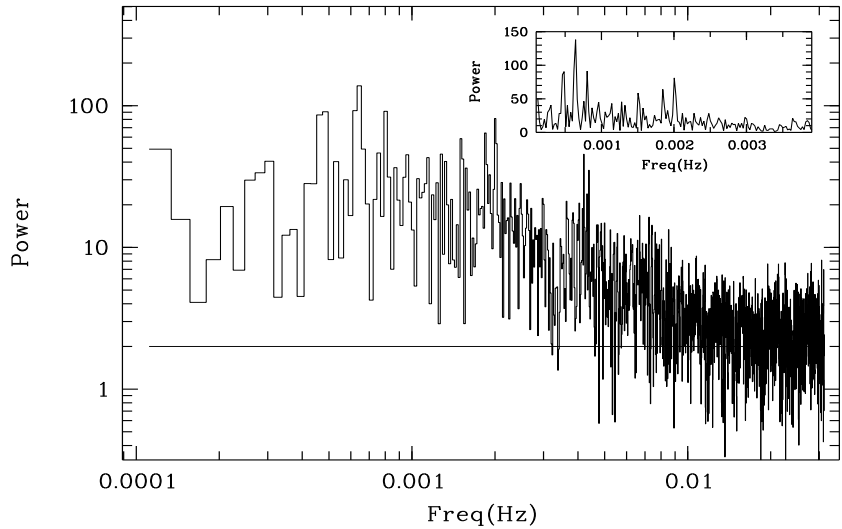}
      \caption{The average power spectrum of TT Ari for SIS0, SIS1, GIS2
            and GIS3 detectors. Data were binned in 16 sec intervals
            for each
            detector. Horizontal line represents
            the counting statistics noise. Inset shows the lower frequency
            part with a linear scale.
              }
%       \label{Fig-1}
\end{figure}

A power spectrum with 16 sec data bins is presented in Fig. 5 which
 has a mean value of 2 at frequencies higher than 10 mHz.
A red noise component (or flicker noise) dominates at
 frequencies between 1mHz and 10 mHz. The noise is saturated
(secondary white noise )
at frequencies lower than 1mHz (see Fig. 5).
%----------------------------------------------------------- Spec_1
\begin{figure}
\vspace{6.5cm}
\includegraphics{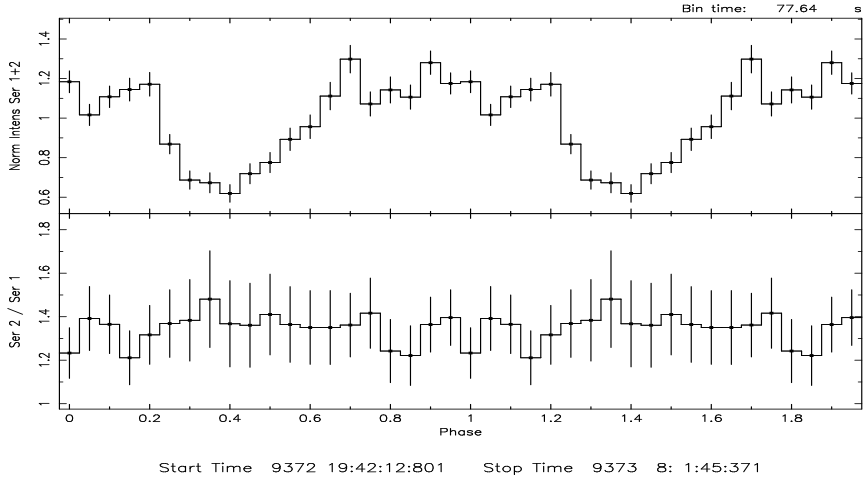}
      \caption{The upper panel is the normalized light curve
            folded at 1553 seconds.
            The lower panel is the
            hardness ratio (2-10 keV/0.2-2 keV) of the ligh curve.
              }
%       \label{Fig-1}
\end{figure}

In order to search for  an oscillation ,
we fitted a constant line to the power at frequencies lower than 1 mHz 
yielding the mean power  $<$Power$>$ = 22.  
In order to see the probability of detection of a significant signal above 
this level, an exponential probability distribution,
Prob(P)=(1/2$\sigma ^{2}$) exp (-P/2$\sigma^{2}$),
 which is essentially a $\chi ^{2}$ distribution for two degrees of 
freedom (Scargle 1982) where $<$Power$>$=2$\sigma^{2}$ =22, is applied.
For  given parameters, the confidence level of the signal 
for the maximum power (P$_{max}$=138) at 0.643 mHZ is
1-N$_{p}$ exp (-P$_{max}$/$<$Power$>$) $\sim$ 0.9 (Leahy et al., 1983),
where N$_{p}$=40 is the number of iteration
frequencies between 1$\times10^{-4}$ and 1$\times10^{-3}$.
This confidence level is close to $\sim 2\sigma$ signal 
detection and the error in the period estimate is 
$\Delta$Period = Period$^{2}$/ T$_{obs} \sim 170$ sec where T is the time span
of the observations.
The symmetric side lobes of the peak frequency are the convolved
window function of the observation gaps. These are seen at ferequencies
 $ 0.643  \pm 0.17$~mHz, where
$ 0.17^{-1}$~mHz$^{-1} \sim 5582$~sec  is close to the
 orbital period of the ASCA satellite around the
earth ($\sim 5700$~sec).
 We also deconvolved the window function
 by making use of the 
"CLEAN" algorithm (Roberts et al., 1987) and verified the $0.643$ mHz
oscillation. In Fig. 6, the light curve folded at this frequency is
presented together with its hardness ratio (2-10 keV/0.2-2 keV).

\section {Discussion }

 We find that  TT Ari is one of the more 
"sophisticated" and interesting sources among the catalysmic binaries. 
 Our basic results are summarised below.

A soft X-ray excess around 
 $\sim 1$~keV  has been seen In magnetic catalysmics binaries 
 (Singh $\&$ Swank 1993, 
 Ishida et al., 1994, Mukai et al., 1994).These systems have spin periods 
$10-10^{3}$~sec. According to the theories of 
 magnetic cataclysmic binaries (such as DQ Her systems),
 most of the X-ray emission 
 is received from the shock regions of the accretion 
 columns near the white dwarf magnetic poles 
 (Patterson 1994). 
 The cooling plasma in the accretion column has both soft 
 $\sim 1$~keV and hard $\sim 9$~keV X-ray 
 components (Singh $\&$ Swank 1993, Ishida et al., 1994). 
 Some spectra show the iron fluorescence line  
 at 6.43 keV which is an indicator of emission from 
 the white dwarf surface (Ishida et al., 1994).
 Although the X-ray spectra of TT Ari
 shows both soft and hard X-ray components,   
 it deviates from the category of   magnetic cataclysmic binaries 
 for at least  two reasons; i) the spin period of the 
 source is not observed, ii) it does not show a 6.43 keV emission line.

EINSTEIN and ROSAT observations have shown that 
the upper limits of the X-ray luminosity from the late type seconderies
 are L$_{x} \sim 10 ^{29}$~erg/sec (Fleming et al., 1989, 
Hempelmann et al., 1995).
 This is much less than the X-ray luminosity of TT Ari 
(L$_{x} \sim 1.8 \times 10^{31}$~erg/sec for a distance of 125 pc).
  This rules out the possible explanation that the X-ray flux comes from
 the corona of the secondary star.
   
 According to Jensen et al., (1983) the X-ray flickering 
 lags the optical flickering by $\sim 1$ minute.  
 This condition implies  the possibility of X-ray 
 emission regions above and below the accretion disk.
 Indeed, the 6.67 keV emission line suggests the possibility of
 emission from an accretion disk corona as observed from
 several X-Ray binaries (Kallman $\&$ White 1989).

 The observed quasi-periodic oscillations 
 in the optical band show  correlation with mass accretion rates 
 (Hollander $\&$ Paradijs 1993) which can be explained 
 by a beat frequency model, as 
 used to explain QPOs in Low Mass X-Ray Binaries (Alpar $\&$ Shaham 1985).
 However, this correlation was not seen in the X-ray band.
 The X-ray flux is 10$\%$ higher  
 in the ASCA observations compared to the  
 ROSAT observations while the peak frequency of flickering 
 oscillations (or quasi-periodic oscillation) decrease  by  
 36$\% $ (see also Baykal et al., 1995).
 This is an another indicator that the X-ray flux 
 is not directly correlated with the mass accretion rate.
  Furthermore, hardness ratios of the 
 folded lightcurve (see Fig. 6) do not show absorption (or 
 hard X-ray excess) at the dips of the lightcurve,
  which implies that the flickering in X-rays is  
 not associated with occultation of blobs.
Our main conclusion is that the X-ray emission may be associated with an
accreation disk corona rather then directly with the accreation column.
 In order to improve our understanding of the flickering oscillations,  
  multiwavelength simultaneous observations 
 of TT Ari are required.                               
                                
\section{Acknowledgements}
We thank the referee Prof. Dr. J.E. Dyson, for a careful reading and 
valuable comments.
A.B acknowledges the National Research
Council for their support. We also thank the {\it ASCA} GOF team 
for the archival data.  
This work is supported by The Scientific and Technical Research Council 
of Turkey,  under High Energy Astrophysics Unit. \\

\end{document}